\documentclass[runningheads]{llncs}

\usepackage[T1]{fontenc}
\usepackage{graphicx}
\usepackage[compatibility=false]{caption}
\usepackage{subcaption}
\usepackage{algorithm}
\usepackage{algpseudocode}
\usepackage{amsmath,amssymb}
\usepackage{float}
\usepackage{hyperref}
\usepackage{amsfonts}
\usepackage{booktabs}
\usepackage{multirow}
\usepackage{esvect}
\hypersetup{
    colorlinks=true,       
    linkcolor=blue,        
    filecolor=magenta,     
    urlcolor=cyan,
}

\begin{document}

\title{Fairness-Aware Profit Maximization using Deep Reinforcement Learning}

\author{Poonam Sharma\inst{1}\orcidID{0009-0009-2608-9414} \and
Sanchit Virdi\inst{2} \and
Suman Banerjee\inst{1}\orcidID{0000-0003-1761-5944}}

\authorrunning{Sharma et al.}

\institute{Indian Institute of Technology Jammu, Jammu \& Kashmir, India\\
\email{\{poonam.sharma,suman.banerjee\}@iitjammu.ac.in}
\and
National Institute of Technology Srinagar, Jammu \& Kashmir, India\\
\email{virdisanchit@gmail.com}}

\maketitle

\begin{abstract}
Given a social network represented as a graph where the nodes are the users and the edges represent the social relations, and a positive integer $k$, how to select $k$ nodes to maximize the influence in the network remains an active area of research. In this paper, we consider a variant of the problem in which network users are associated with two parameters: a benefit value and a cost. A fixed budget is given, and the network is partitioned into communities. The task is to select a subset of users (the seed set) within the budget so that their initial activation maximizes the earned profit, while ensuring that each community realizes at least a minimum fraction of its total benefit under a maximin fairness criterion. For any seed set, the earned benefit is defined as the sum of the benefit values of the users influenced by the seed set, and the profit is defined as the difference between the earned benefit and the total cost. Formally, we call this the \emph{Fairness-Aware Profit Maximization Problem}. We propose a Deep Reinforcement Learning-based approach for solving it: we first model the problem as a \emph{Markov Decision Process} and subsequently propose a \emph{Deep $Q$-Learning Algorithm}. The proposed solution has been implemented and tested on real-world social network datasets. From the reported results, we observed that the proposed approach yields a seed set whose initial activation produces up to $10$ times more profit than the baseline methods. The implementation of our methodology is available at \url{https://github.com/PoonamSharma-PY/DRL_FPM.git}.
\end{abstract}
\section{Introduction}
A social network can be defined as an interconnected structure among a group of human agents, that is formed for social interactions \cite{borgatti2024analyzing}. Due to the advent of social networks and mobile handheld devices, the use of online social networks has become ubiquitous in day-to-day life \cite{azzaakiyyah2023impact}. The commercial houses use these online social networks for campaigning purpose in the following way. They distribute a limited number of products to influential people in the network with the hope that many of them will like the product and share positive feedback as a social media post. As they are influential people, they have many followers and many of them will actually like the post and share it \cite{rachmad2022social}. This process will continue, and at the end of the diffusion process, a significant number of users will be influenced. The key computational problem in this case is given a social network and a positive integer $k$, choose a $k$-sized subset as a seed set to maximize the influence. This problem has been referred to as the Influence Maximization Problem in the literature. There exists a significant amount of literature on this problem \cite{banerjee2020survey,li2018influence}.  
\par In recent times, a variant of this problem has been introduced where the users of the network are associated with two positive integers, which we call the cost and benefit associated with the user. The cost of the user signifies the amount of incentive that needs to be paid if the user is chosen as seed user, and the benefit of the user signifies the amount of benefit that can be earned if the user becomes influenced. For any subset of user, the earned benefit of the subset is defined as the sum of the benefit values of the influenced users in the network when the diffusion starts from this subset. The profit of the seed set is defined as the difference between the benefit earned by the seed set and its cost. In the context of viral marketing, the goal will be to maximize profit \cite{tang2017profit,chen2020random}. Hence, the problem becomes, for a given budget how to select the seed nodes effectively such that the earned profit by the seed set gets maximized. There are several solution methodologies in the literature. However, to the best of our knowledge, the notion of the \emph{fairness} has not been considered in the existing studies.
\par In practice, a social network consists of a number of communities, and a community in a social network is defined as a subset of the users such that they are densely connected among themselves and sparsely connected among other parts of the network \cite{singh2024social}. In this setup, it will be good to ensure fairness due to the following reason. If the ultimate objective is to maximize the total earned profit, then it may so happen that a significant amount of the earned profit is from a very few communities. This scenario signifies that in the rest of the communities, the influence is very limited. This is, of course, not a very desirable situation because in that case, many people will be unaware of the product. While deep reinforcement learning has been applied to fairness-aware influence maximization~\cite{Saxena2025DQ4FairIMFI}, no comparable approach exists for the profit-aware setting, where node-level costs and benefits induce a budget constraint and change the structure of the seed selection problem. This motivates us to define the \textsc{Fairness-Aware Profit Maximization} Problem, and we have proposed a Deep $Q$-Learning Algorithm to solve it. In particular, we make the following contributions in this paper:
\begin{itemize}
    \item We have introduced and studied the Fairness-Aware Profit Maximization Problem, for which the available literature is limited.
    \item  We have proposed a Deep $Q$-Learning Algorithm (RL4FPM) for solving this problem.
    \item  We have experimented with real-world social network datasets and compared our proposed approach with the existing methodologies.
\end{itemize}
The rest of the paper has been organized as follows. Section \ref{Sec:PD} describes the background information and formally defines the problem. Section \ref{Sec:Problem} describes the formulation of our problem. The proposed solution methodology has been described in Section \ref{Sec:Methodology}. The experimental evaluation of the proposed approach has been described in Section \ref{Sec:Experiments}. Finally, Section \ref{Sec:CFD} concludes our study and poses some future research questions. 

\section{Background and Problem Definition} \label{Sec:PD}
This section describes the background of our study and formally defines our problem. Initially, we start by describing the notion of social networks and its diffusion phenomenon.

\subsection{Social Network and Influence Propagation}
We model a social network as an edge-weighted, (un)directed graph $G(V,E,p)$, where $V$ is the set of users, $E$ is the set of social relations, and $p:E \to (0,1]$ maps each edge to an influence probability. We adopt the \emph{Independent Cascade (IC) model} for diffusion: starting from a seed set $S \subseteq V$, an influenced node has one chance to activate each of its uninfluenced neighbors, with success probability equal to the edge weight. The diffusion proceeds in discrete steps until no further activation occurs. The expected number of nodes influenced by $S$ is denoted $\sigma(S)$, with $\sigma: 2^V \to \mathbb{R}_0^+$.

\subsection{Profit Maximization Problem}
Each user has an associated cost and benefit, given by functions $\mathcal{C}: V \to \mathbb{R}^+$ and $b: V \to \mathbb{R}^+$. For any subset $S \subseteq V$, the total selection cost is $\mathcal{C}(S) = \sum_{u \in S} \mathcal{C}(u)$. If $S$ is the seed set, let $I(S)$ denote the set of influenced nodes. The earned benefit is $\beta(S) = \sum_{u \in I(S)} b(u)$, and the profit is $\Pi(S) = \beta(S) - \mathcal{C}(S)$. The Profit Maximization Problem is to select a seed set within a fixed budget $B$ that maximizes the profit:
\begin{equation}
    S^{OPT} \longleftarrow \underset{S \subseteq V,\ \mathcal{C}(S) \leq B}{\arg\max} \Pi(S).
\end{equation}
Setting $b(v) = n$ and $\mathcal{C}(v) = 1$ for all $v$, with budget $B = k$, reduces $k$-cardinality Influence Maximization to Profit Maximization (the constant $-|S|$ term is dominated by $n \cdot |I(S)|$). Since $k$-IM is NP-hard~\cite{kempe2003maximizing}, so is Profit Maximization.

\subsection{Network Embedding}
A network embedding maps each node to a low-dimensional vector, $h: V \to \mathbb{R}^d$, with $h_v \in \mathbb{R}^d$ for node $v$. We adopt Structure2Vec~\cite{215331}, specifically the DE-MF variant~\cite{dai2016discriminative}, which iteratively refines $h_v$ by aggregating its neighbors' embeddings together with the node's own features (cost and benefit). The resulting embeddings serve as input features to the Q-network described later.

\subsection{Markov Decision Process}
A Markov Decision Process (MDP) is a tuple $(\mathcal{S}, \mathcal{A}, P, R, \gamma)$, where $\mathcal{S}$ is the state space, $\mathcal{A}$ the action space, $P$ the transition function, $R$ the reward function, and $\gamma \in [0,1)$ the discount factor. The action-value function under policy $\pi$ is
\begin{equation} \label{Eq:RL_1}
Q^{\pi}(s,a) = \mathbb{E}_{\pi}[G_t \mid S_t = s, A_t = a],
\end{equation}
and the goal of reinforcement learning is to learn an optimal policy $\pi^*$ that maximizes the expected total reward.

\section{Problem Formulation} \label{Sec:Problem}
We consider that the given social network is partitioned into $\ell$ disjoint communities denoted by $C = \{C_1, C_2, \dots, C_i, \dots, C_\ell\}$. Hence, each user of the network belongs to exactly one community. Given a seed set $S \subseteq V$ and the per-node benefit $b(v)$, let $\eta_{C_i}(G,S)$ denote the \emph{community benefit ratio} of $C_i$, defined as the fraction of $C_i$'s total benefit that is realized under the seed set $S$:
\begin{equation} \label{Eq:CommunityCoverage}
\eta_{C_i}(G,S) \;=\; \frac{\sum_{v \in C_i \cap I(S)} b(v)}{\sum_{v \in C_i} b(v)},
\end{equation}
where $I(S)$ is the set of nodes influenced by $S$ under the diffusion process. To quantify fairness in influence spread, we use a benefit-weighted maximin fairness criterion, which ensures that each community earns at least a minimum fraction of its total realizable benefit. This can be computed as shown in Equation \ref{Eq:Fairness}.
\begin{equation} \label{Eq:Fairness}
f(G,S) \;=\; \min_{\forall C_i \in C} \; \eta_{C_i}(G,S).
\end{equation}
Maximin fairness ensures that the least-served community receives more benefit from the newly selected seed node, promoting equitable benefit realization across all groups.

\subsection{Problem Statement}
In this paper, we study the problem of selecting a seed set within a given budget that maximizes the overall profit while ensuring maximal fairness in benefit realization across communities, measured using the maximin fairness criterion. We refer to this as the \emph{Fairness-Aware Profit Maximization Problem}. Formally, it is stated as follows:
\begin{align}
    & \max_{S \subseteq V} \; \Pi(G,S) \label{Eq:Objective} \\
    \text{subject to} \quad & \mathcal{C}(S) \leq B, \label{Eq:BudgetConstraint} \\
    & f(G,S) \geq \tau, \label{Eq:FairnessConstraint}
\end{align}
where $\Pi(G,S)$ is the net profit obtained from the seed set $S$, $\mathcal{C}(S)$ is the total selection cost, $B$ is the allocated budget, and $\tau$ is the minimum acceptable community benefit ratio under the maximin criterion of Equation~\ref{Eq:Fairness}. The first constraint ensures that the cost of the seed set is within the allocated budget, and the second ensures that the worst-served community receives at least a $\tau$ fraction of its total realizable benefit. If multiple seed sets yield the same maximum profit, any set that additionally maximizes overall outreach can be chosen. Next, we model the Fairness-Aware Profit Maximization problem using Reinforcement Learning (RL).

\subsection{Formulation using Reinforcement Learning (RL)}
Here, we describe different attributes of the Markov Decision Process. First, we start by describing the notion of a timestep. \\
\begin{itemize}
    \item \textbf{Timestep:} In our study, we select seed nodes in a round-wise manner. Each round of seed selection is modeled as a discrete timestep $t$, with one node selected per round.
The time horizon is indicated by $t = 1, \dots, T$ with $T = k$, where $k$ is the maximum number of seed nodes allowed. Hence, $k \leq \frac{B}{\mathcal{C}_{\min}}$, where $\mathcal{C}_{\min}$ denotes the minimum selection cost, i.e., $\mathcal{C}_{\min} = \underset{u \in V}{\min} \ \mathcal{C}(u)$. Thus, the agent selects one node at each timestep until the seed set reaches size $k$.
    \item \textbf{State:} At timestep $t$, the state $S_t$ is represented by the tuple $S_t = (G, C, X_t)$, where $G = (V, E)$ is a graph sampled from the pool of graphs $\mathcal{G}$, $C$ is the set of communities in the network, and $X_t \in \{0,1\}^{|V|}$ is a binary vector indicating which nodes have been selected as seed nodes up to time $t$. $X_t^v = 1$ if a node $v$ is selected as a seed till the timestep $t$, and $0$ otherwise. Initially, no node is selected, and therefore, $X_1^v = 0$ for all $v \in V$. While the MDP state $S_t = (G, C, X_t)$ formally includes the binary selection vector $X_t$, the Q-network's input is a compact encoding that summarizes $X_t$ by its size $|X_t|$ together with a Structure2Vec embedding of the candidate node.

    \item \textbf{Action:} At each timestep $t$, the agent selects a node to add to the seed set, and its cost is subtracted from the allocated budget. This is represented by a one-hot vector $a_t \in \{0,1\}^{|V|}$  with one non-zero entry indicating the chosen node i.e., $\sum_{v=1}^{|V|} a_t^v = 1$. The action space at the state $S_t$ is constrained both by selection $X_t^v = 0$ and by feasibility under the remaining budget $\mathcal{C}(v) \leq B_{rem,t}$.
    
    \item \textbf{State-action-transition probability:} When a node is selected at timestep $t$, it is added to the seed set and the state is updated accordingly as shown in Equation \ref{Eq:State_1} and \ref{Eq:State_2}.
    \begin{equation} \label{Eq:State_1}
      X_{t+1} = X_t + a_t, \quad t = 1, \dots, T.  
    \end{equation}
    \begin{equation} \label{Eq:State_2}
    B_{\text{rem}, t+1} = B_{\text{rem}, t} - \mathcal{C}(v), \quad t = 1, \dots, T,
    \end{equation}
    
    \item \textbf{Reward:} The total reward at the end of an episode is defined as the total profit in the network $G$ resulting from the selected seed set $S$. This formulation implies that the agent would receive a reward only at the terminal time step $T$, not during intermediate steps. This sparsity of rewards can significantly hinder the agent’s learning efficiency. To address this, we define an immediate reward at each time step using the marginal profit of the selected node, and this can be mathematically defined in Equation \ref{Eq:reward}
 
    \begin{equation} \label{Eq:reward}
r(G,X_t,a_t) = \Pi\bigl(G,\, S(X_t) \cup \{v\}\bigr) - \Pi\bigl(G,\, S(X_t)\bigr),
\end{equation}
where $v$ is the node selected at time $t$ ($v \in V$ and $a_t^v = 1$), and $S(X_t) = \{u \in V \mid X_t^u = 1\}$ denotes the seed set induced by $X_t$. With slight abuse of notation, we write $S$ for $S(X_t)$ in the remainder of the paper when the timestep is clear from context.
    \item \textbf{Enhancing Fairness:} To incorporate fairness, we add a maximin term to the reward, weighted by $\varphi \geq 0$:
    \begin{equation} \label{Eq:E_Fairness}
    R(G,S) = \Pi(G,S) + \varphi \cdot f(G,S).
    \end{equation}
    The corresponding per-step marginal reward is
    \begin{equation} \label{Eq:Agree_Fairness}
r(G,X_t,a_t) = \bigl(\Pi(G,\, S(X_t) \cup \{v\}) - \Pi(G,\, S(X_t))\bigr) + \varphi \bigl(f(G,\, S(X_t) \cup \{v\}) - f(G,\, S(X_t))\bigr).
\end{equation}
    In practice, $\Pi$ and $f$ are estimated from a single Independent Cascade rollout per step; the resulting stochasticity is absorbed by replay-buffer averaging at the gradient update.
\end{itemize}

\section{Proposed Solution Approach} \label{Sec:Methodology}
In this section, we propose \textbf{RL4FPM}, a deep Q-learning approach for budget-constrained profit maximization that learns a seed-selection policy generalizing across graph instances. Algorithm~\ref{Algorithm1:DRLFairPM} trains a Q-network $Q_\theta$ that scores the marginal value of adding any candidate node under a budget constraint. Each episode samples a fresh subgraph $G$ from the training pool; Structure2Vec produces node embeddings $\{h_v\}_{v\in V}$ (collectively denoted $h$ in the algorithm), and the agent constructs a seed set one node at a time via an $\varepsilon$-greedy policy over budget-feasible candidates. The reward at each step is the marginal increment of $R = \Pi^{\text{IC}} + \varphi F$, where $\Pi^{\text{IC}}$ is a single-sample IC benefit and $F$ is the worst-community fairness term. Transitions are stored in a fixed-capacity replay buffer $\mathcal{M}$, and every $T_u$ episodes, a uniform minibatch is drawn to perform a single gradient step on the mean squared one-step Bellman error. Because each episode samples a different subgraph, $\mathcal{M}$ is cross-graph, which trains $Q_\theta$ toward a graph-conditional value function and allows reuse on unseen test subgraphs.

\begin{algorithm}[!htb]
\caption{RL4FPM: Deep Q-Learning for Fairness-Aware Profit Maximization}
\label{Algorithm1:DRLFairPM}
\scriptsize
\begin{algorithmic}[1]
\Require Budget $B$; episodes $E_p$; learning rate $\alpha$; discount $\gamma$; $\varepsilon$-schedule $(\varepsilon_0,\varepsilon_{\min},\rho)$; replay capacity $N$; batch size $K$; update period $T_u$; min.\ cost $\mathcal{C}_{\min}$; fairness weight $\varphi$
\Ensure Trained Q-network $Q_\theta$
\State Initialize $Q_\theta$ with random weights $\theta$;\quad $\mathcal{M}\gets\emptyset$;\quad $\varepsilon\gets\varepsilon_0$
\For{$e = 1$ to $E_p$}
    \State Sample training subgraph $G=(V,E)$; load $b(\cdot),\mathcal{C}(\cdot),C$
    \State $h \gets \textsc{Structure2Vec}(G,\mathcal{C},b)$
    \State $X\gets\emptyset$;\ \ $Z\gets\emptyset$;\ \ $B_\text{rem}\gets B$;\ \ $R_\text{prev}\gets 0$
    \While{$B_\text{rem}\ge \mathcal{C}_{\min}$}
        
        \State $a_t \gets \textsc{$\varepsilon$-Greedy}(Q_\theta,G,X,Z,\varepsilon,h,B_\text{rem},\mathcal{C})$
        \If{$a_t=\emptyset$} \textbf{break}
        \EndIf
        \If{$\mathcal{C}(a_t)>B_\text{rem}$}
            \State $Z\gets Z\cup\{a_t\}$;\ \ \textbf{continue}
        \EndIf
        \State $B_\text{rem}\gets B_\text{rem}-\mathcal{C}(a_t)$;\quad $X'\gets X\cup\{a_t\}$
        \State $R_t \gets \textsc{ProfitFairnessReward}(G,X',b,\varphi)$
        \State $r_t \gets R_t-R_\text{prev}$;\quad $R_\text{prev}\gets R_t$
        \State $\text{term}\gets (B_\text{rem}\le 0)$
        \State Store $(G,X,a_t,r_t,X',\text{term},h)$ in $\mathcal{M}$
        \State $X\gets X'$
    \EndWhile
    \If{$e \bmod T_u = 0$ \textbf{and} $|\mathcal{M}|\ge K$}
        \State Sample minibatch $\mathcal{B}\subset\mathcal{M}$ of size $K$;\quad $L\gets 0$
        \For{each $(G,X,a,r,X',\text{term},h)\in\mathcal{B}$}
        \State $s\gets\textsc{EncodeState}(G,X,a,h,B_\text{rem},\mathcal{C})$
        \If{$\text{term}$}
            \State $y\gets r$
        \Else
            \State $y \gets r + \gamma \max_{v \in V \setminus X'} Q_\theta\!\left(\textsc{EncodeState}(G, X', v, h, B_\text{rem}, \mathcal{C})\right)$
        \EndIf
            \State $L\gets L + (y-Q_\theta(s))^2$
        \EndFor
        \State $\theta\gets\theta-\alpha\nabla_\theta(L/K)$
    \EndIf
    \State $\varepsilon\gets\max(\varepsilon\cdot\rho,\,\varepsilon_{\min})$
\EndFor
\State \Return $Q_\theta$
\end{algorithmic}
\end{algorithm}

\paragraph{\textbf{Time complexity of RL4FPM}.}
Gradient updates are performed every $T_u$ episodes, and the IC reward uses a single sample per step. Under these settings, the training time complexity for a pool of graphs is
\[
O\bigl(HE_p\,(|V|\cdot|\theta| + |E|) \;+\; E_p T_{\text{emb}}\,|E|\,d_{\text{emb}}\bigr).
\]
Here, $H$ is the episode horizon (bounded by $B/\mathcal{C}_{\min}$), and $E_p$ is the number of training episodes. $|V|$ and $|E|$ are the node and edge counts of the sampled subgraph, and $|\theta|$ is the number of weights in the Q-network $Q_\theta$. $T_{\text{emb}}$ is the number of Structure2Vec iterations, and $d_{\text{emb}}$ is its embedding dimension. The first term covers $\varepsilon$-greedy action selection over $|V|$ candidates per step, along with the single-sample IC rollout. The second term covers the per-episode Structure2Vec cost. The gradient-update cost is amortized into the action-selection term and is not shown separately.

In inference, the trained policy is run greedily on test graphs. The reward rollout is no longer needed, so the per-instance cost reduces to
\[
O\bigl(T_{\text{emb}}\,|E|\,d_{\text{emb}} \;+\; H\,|V|\cdot|\theta|\bigr).
\]
A classical Monte Carlo greedy profit-IM solver instead costs $O(H\cdot|V|\cdot m\cdot|E|)$ per instance, where $m$ is the number of Monte Carlo simulations. The $m\cdot|E|$ factor is paid once per candidate by the greedy method. RL4FPM replaces this with a single $O(|V|\cdot|\theta|)$ Q-network evaluation per candidate, and the IC-rollout cost is paid only during training.

\section{Experimental Details} \label{Sec:Experiments}

\subsection{Dataset Description}

The experiments are performed on synthetic and real-world networks. All datasets are partitioned into majority and minority communities at a $20{:}80$ ratio. The real-world networks are downloaded from \href{https://snap.stanford.edu/index.html} {SNAP: Stanford Network Analysis Project}. These datasets are listed in Table \ref{Table1:Datasets}. 

\begin{table}[htbp]
\centering
\caption{Basic statistics of the datasets used in our experiments.}
\label{Table1:Datasets}
\scriptsize
\setlength{\tabcolsep}{6pt}
\renewcommand{\arraystretch}{1.2}
\begin{tabular}{@{}llrrrcc@{}}
\toprule
\multirow{2}{*}{\textbf{Dataset}} & \multirow{2}{*}{\textbf{Type}} & \multicolumn{2}{c}{\textbf{Source Network}} & \multicolumn{3}{c}{\textbf{Sampled Instances}} \\
\cmidrule(lr){3-4} \cmidrule(lr){5-7}
 & & $|V|$ & $|E|$ & Nodes & Train & Test \\
\midrule
Email-Eu-Core & Directed   & $1{,}005$  & $25{,}571$  & $500$   & $12$ & $8$  \\
H-BA2k        & Undirected & $30{,}000$ & $119{,}990$ & $2{,}000$ & $50$ & $10$ \\
Wiki-Vote     & Directed   & $7{,}115$  & $103{,}689$ & $1{,}000$ & $12$ & $8$  \\
\bottomrule
\end{tabular}
\end{table}

\subsection{Experimental Setup}
\paragraph{\textbf{RL Parameters}} We trained \textbf{RL4FPM} for $E_p = 720$ episodes using the Adam optimizer with learning rate $\alpha = 0.001$ and discount factor $\gamma = 0.95$. Exploration followed an $\varepsilon$-greedy schedule, annealed multiplicatively by $\rho = 0.995$ per episode from $\varepsilon_0 = 1.0$ toward a floor of $\varepsilon_{\min} = 0.05$. The updates were performed on minibatches of size $K = 32$ drawn uniformly from a FIFO replay buffer of capacity $N = 10{,}000$, with one gradient step taken every $T_u = 2$ episodes. The Q-network was a feed-forward architecture with hidden dimension $128$ and ReLU activations, taking as input a $67$-dimensional state vector that concatenates a $d_{\text{emb}} = 64$ Structure2Vec embedding (generated with $T_{\text{emb}} = 3$ message-passing iterations) with the candidate node's cost, the remaining budget, and the current seed-set size. The reward combined the realized benefit earned from the influence of a single Independent Cascade rollout with a worst-community fairness term weighted by $\varphi = 1.0$.

\paragraph{\textbf{Influence Probability}} We use two settings: (i) Uniform, with $p(u_i,u_j) = 0.1$ for all $(u_i,u_j) \in E$, and (ii) Trivalency, with each edge's probability drawn uniformly at random from $\{0.1, 0.01, 0.001\}$.

\paragraph{\textbf{Budget(B)}}: We have experimented with the $6$ different budget values: $500$, $1000$, $1500$, $2000$, $2500$ and $3000$.

\paragraph{\textbf{Evaluation Metric}}: We evaluate the profit metric for all the algorithms over a range of budget values, averaged over $m = 1{,}000$ Monte Carlo simulations.

\paragraph{\textbf{Environment Setup}} All experiments were conducted on a Linux workstation with a 32-core Intel processor (3.2 GHz) and 64 GiB of memory. All algorithms are implemented in a Python 3.13.5 environment, integrated with the NetworkX 3.5 and torch 2.9.0 libraries.

\subsection{Baseline Methods}

We compare our approach with the following baselines:
\begin{itemize}
\item \textbf{Random}: Selects the seed nodes randomly within the given budget. 
\item \textbf{Simple-PageRank}: Selects top-k nodes based on the Pagerank centrality.
\item \textbf{Parity} \cite{10.1145/3366423.3380275}: Parity seeding is a fairness-aware strategy that modifies the selection thresholds for different groups to balance representation in the seed set. Overall, it selects nodes based on degree, ensuring that the group proportions of the seed set match those of the entire population.
\item \textbf{High Degree}: Selects the top-k nodes having the highest degree and within the given budget.
\item \textbf{Fair-PageRank}: The parity-based seeding is applied on PageRank centrality instead of degree, ensuring that seed nodes are fairly selected from all communities.
\item \textbf{Crosswalk} \cite{pantea2023re}: Crosswalk generates fairness-aware network embeddings and selects the most centrally located nodes as seed nodes using the k-medoids clustering method.

\end{itemize}

\subsection{Results and Discussion}
\begin{figure}[htbp]
  \centering
  \includegraphics[width=\linewidth, height=0.50\textheight, keepaspectratio]{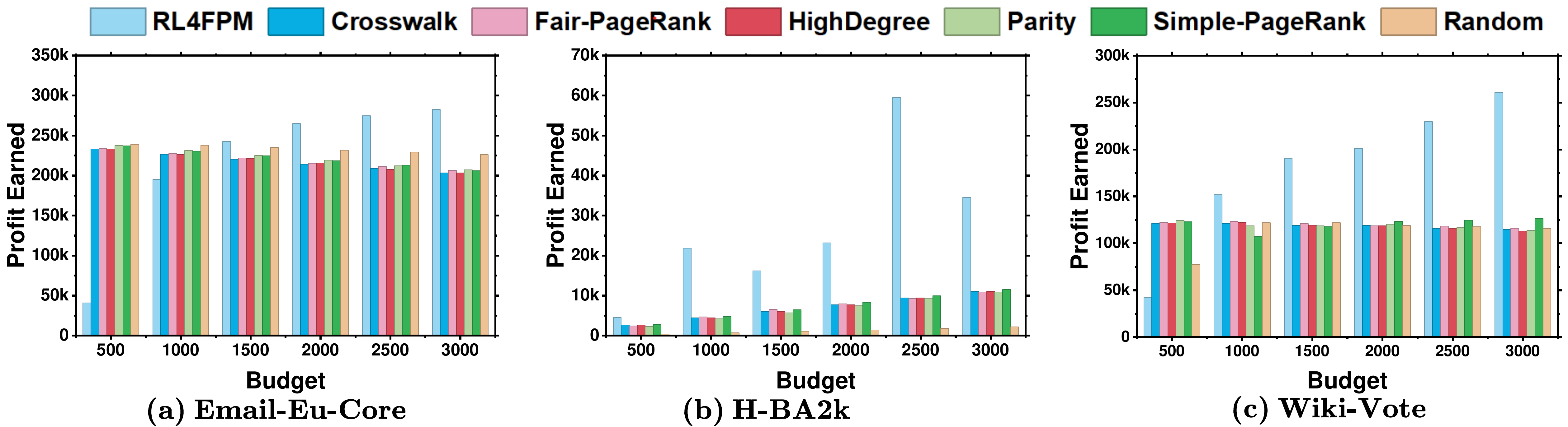}
  \caption{Budget Vs. Profit Plots for Uniform Probability settings}
  \label{Plot1:ProfitUniform}
\end{figure}

\begin{figure}[htbp]
  \centering
  \includegraphics[width=\linewidth, height=0.50\textheight, keepaspectratio]{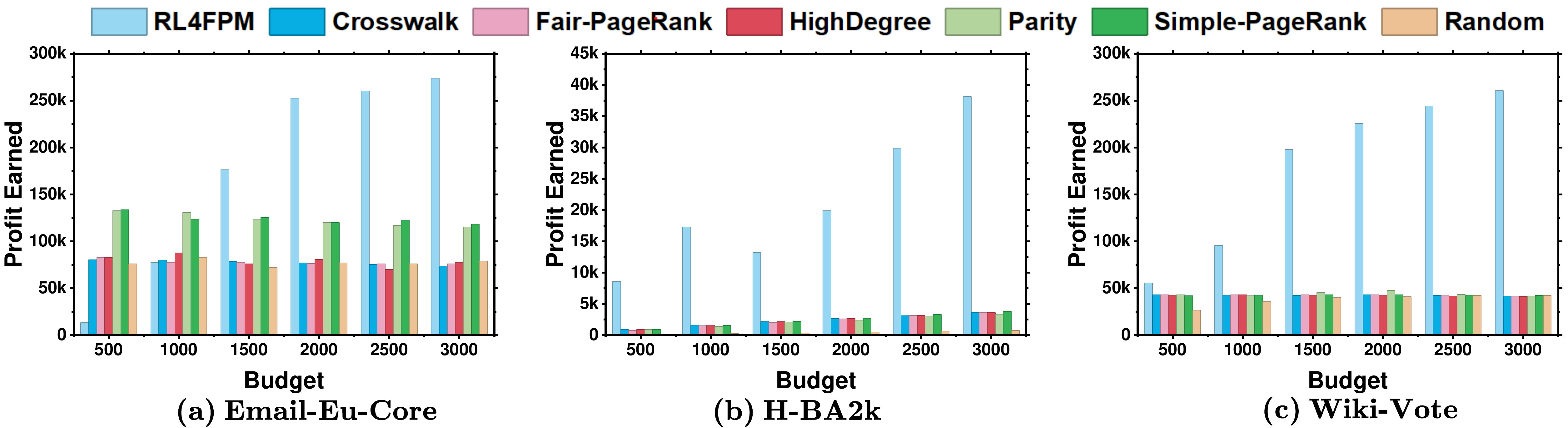}
  \caption{Budget Vs. Profit Plots for Trivalency Probability settings}
  \label{Plot2:ProfitTrivalency}
\end{figure}

In this section, we will discuss the results of our experiments. Our discussion will be through the following analysis.
\paragraph{\textbf{Impact of Budget on Profit Earned}}
We study profit earned by RL4FPM and the baselines across budgets, starting with the uniform probability setting. On Email-Eu-Core
(Figure \ref{Plot1:ProfitUniform}(a)), Random is the strongest baseline across all budgets. RL4FPM achieves improvements of $3\%$, $14\%$, $20\%$, and $25\%$ over Random at budgets $1500$, $2000$, $2500$, and $3000$. On H-BA2k (Figure \ref{Plot1:ProfitUniform}(b)), RL4FPM achieves maximum profit at every budget. Simple-PageRank is the strongest baseline (Fair-PageRank at $1500$). The profit curve is non-monotonic. Improvements range from $60\%$ vs Simple-PageRank at $500$ to $501\%$ at $2500$. The volatility points to training variance. On Wiki-Vote (Figure \ref{Plot1:ProfitUniform}(c)), RL4FPM achieves maximum profit at all but one budget. The strongest baseline shifts: Parity at $500$, Fair-PageRank at $1000$, Random at $1500$, Simple-PageRank from $2000$. The single loss is at $500$ ($-66\%$ vs Parity). Improvements are $23\%$ vs Fair-PageRank at $1000$, $56\%$ vs Random at $1500$, then $63\%$, $84\%$, and $106\%$ vs Simple-PageRank at $2000$, $2500$, and $3000$.
For the trivalency probability setting: On Email-Eu-Core (Figure \ref{Plot2:ProfitTrivalency}(a)), RL4FPM achieves higher profit on most budgets. Simple-PageRank is the strongest baseline (Parity at $1000$). RL4FPM loses at $500$ ($-90\%$ vs Simple-PageRank) but climbs to $132\%$ over Simple-PageRank at $3000$. On H-BA2k (Figure \ref{Plot2:ProfitTrivalency}(b)), RL4FPM achieves maximum profit at every budget. This is the strongest result. High Degree is the strongest baseline at $500$ and $1000$, and Simple-PageRank from $1500$. Improvements are $836\%$ and $978\%$ vs High Degree at $500$ and $1000$, and $897\%$ vs Simple-PageRank at $3000$. On Wiki-Vote (Figure \ref{Plot2:ProfitTrivalency}(c)), RL4FPM achieves maximum profit at every budget with monotonic dominance. The strongest baseline shifts: Crosswalk at $500$, Fair-PageRank at $1000$, Parity at $1500$, $2000$, and $2500$, and Random at $3000$. RL4FPM climbs from $55709$ to $260629$ while baselines stay flat around $42000$ to $45000$. Improvements grow from $29\%$ vs Crosswalk at $500$ to $513\%$ vs Random at $3000$.

We restrict the seed set size and execution time analysis below to the uniform probability setting only, as the trivalency setting yields qualitatively similar patterns and is omitted for space constraints.

\begin{figure}[htbp]
  \centering
  \includegraphics[width=\linewidth, height=0.50\textheight, keepaspectratio]{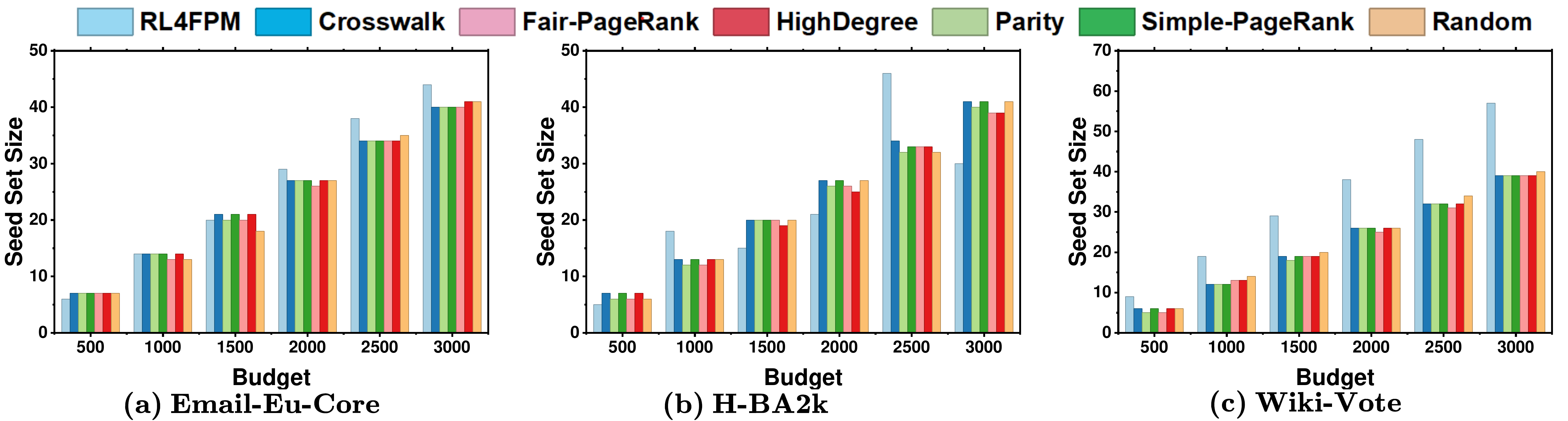}
    \caption{Budget Vs. Seed Set Size Plots for Uniform Probability settings}
    \label{Plot3:SeedSetSize}
\end{figure}

\paragraph{\textbf{Impact of Budget on Seed Set Size}}
A natural question is whether RL4FPM earns more profit by selecting more seeds or better seeds. Figure \ref{Plot3:SeedSetSize} shows that the answer is dataset-dependent. On Email-Eu-Core, RL4FPM picks similar seed counts to baselines but achieves $3\%$ to $25\%$ higher profit, indicating smarter selection. On H-BA2k, it picks fewer seeds (e.g., 5 vs 7 at budget $500$; 30 vs 39 at budget $3000$) yet earns up to $200\%$ more profit, showing that quality dominates quantity. On Wiki-Vote, it picks about $50\%$ more seeds at roughly $52$ vs $78$ cost per seed, trading individual cost for coverage, with profit improvements growing from $23\%$ to $106\%$ over the strongest baseline. RL4FPM thus adapts its seed-set composition to the graph structure: fewer-but-better on H-BA2k, more-but-cheaper on Wiki-Vote, smarter-at-same-count on Email-Eu-Core.

\begin{figure}[htbp]
\centering
  \centering
  \includegraphics[width=\linewidth, height=0.50\textheight, keepaspectratio]{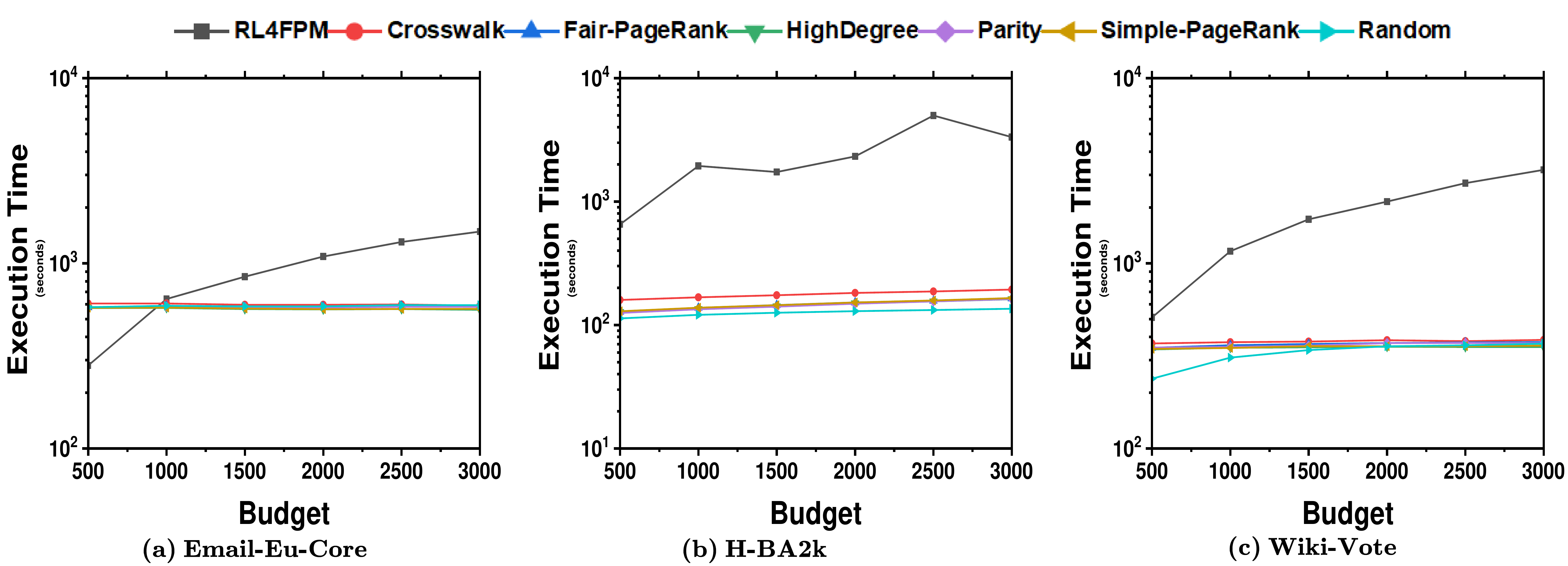}
\caption{Budget Vs. Execution Time (in Seconds) Plots for Uniform Probability settings}
\label{Plot4:ExecutionTime}
\end{figure}

\paragraph{\textbf{Computational performance of the Algorithms}}
Figure \ref{Plot4:ExecutionTime} shows execution time under the uniform setting. Baselines run within about $50$ seconds of each other per dataset, while RL4FPM is the slowest in nearly every configuration. The slowdown ranges from $2.1\times$ (Wiki-Vote, budget $500$) to $37.6\times$ (H-BA2k, budget $2500$). RL4FPM is also the most profitable from budget $1000$ onward on Wiki-Vote and Email-Eu-Core, and at every budget on H-BA2k. The method trades execution time for profit, with the trade-off most pronounced on the largest network.

\section{Concluding Remarks and Future Research Directions} \label{Sec:CFD}
In this paper, we have introduced and studied the Fairness-Aware Profit Maximization Problem, where the goal is to select a seed set within an allocated budget that maximizes the earned profit while ensuring that each community's realized benefit ratio meets a minimum threshold under the maximin fairness criterion. To solve this problem, we proposed RL4FPM, a deep Q-learning approach that combines Structure2Vec embeddings with an $\varepsilon$-greedy seed-selection policy trained over a pool of subgraphs, enabling generalization to unseen networks without retraining. We evaluated RL4FPM on one synthetic and two real-world benchmark networks against six fairness-agnostic and fairness-aware baselines under both uniform and trivalency influence-probability settings. The results show that RL4FPM achieves substantially higher profit than all baselines at moderate-to-large budgets. As future work, we plan to replace the Monte Carlo Independent Cascade rollout with a learned graph neural network surrogate, which would reduce the per-step reward-evaluation cost and allow the method to scale to substantially larger networks.

\bibliographystyle{splncs04}
\bibliography{paper}

\end{document}